\documentclass[11pt]{article}
\usepackage{array}
\usepackage[fleqn]{amsmath}
\usepackage{amssymb}
\usepackage[pdftex]{graphicx}        

\begin{document}

\begin{center}
{\Large Pearson $\chi^2$-divergence Approach to Gaussian Mixture Reduction \\
and its Application to Gaussian-sum Filter and Smoother}

\vspace{7mm}
{\large Genshiro Kitagawa

Mathematics and Informatics Center,
The University of Tokyo\\
7-3-1 Hongo, Bunkyo-ku, Tokyo 113-8656, JAPAN
}

\end{center}

\vspace{10mm}

\begin{center}{\bf\large Abstract}\end{center}

\begin{quote}
The Gaussian mixture distribution is important in various statistical problems.
In particular it is used in the Gaussian-sum filter and smoother for
linear state-space model with non-Gaussian noise inputs.
However, for this method to be practical, an efficient method of reducing the
number of Gaussian components is necessary.
In this paper, we show that a closed form expression of Pearson $\chi^{2}$-divergence
can be obtained and it can apply to the determination of 
the pair of two Gaussian components in sequential reduction of Gaussian components.
By numerical examples for one dimensional and two dimensional distribution models,
it will be shown that in most cases the proposed criterion performed
almost equally as the Kullback-Libler divergence, for which computationally 
costly numerical integration is necessary.
Application to Gaussian-sum filtering and smoothing is also shown.

\end{quote}

\vspace{5mm}
\noindent Divergence, Pearson $\chi^2$-divergence, Gaussian-sum filter.

\section{Introduction}
\label{introduction}

Reduction of the number of components in Gaussian mixture distribution is important
in various field of statistical problems, data fusion, pattern recognition, supervised learning of multimedia and target tracking\cite{Salmond 1990},\cite{West 1993}.
As an example, consider a linear state space model
\begin{eqnarray}
x_n &=& F_nx_{n-1} + G_nv_n \nonumber \\
y_n &=& H_nx_n + w_n,
\end{eqnarray}
where the system noise $v_n$ and the observation noise $w_n$
are distributed according to a mixture of several Gaussian components:
\begin{eqnarray}
 p(v_n) &=& \sum_{i=1}^{q} \alpha_i \varphi(v_n|\mu_v,Q_i) \nonumber\\
 p(w_n) &=& \sum_{j=1}^{r} \beta_j \varphi(w_n|\mu_w,R_j).
\end{eqnarray}
$q$ and $r$ are the number of Gaussian components of
$p(v)$ and $p(w)$, respectively, and $\varphi(x|\mu ,V)$ denotes the Gaussian density with
mean vector $\mu$ and the variance covariance matrix $V$.

Here assume that $Y_n$ denotes the set of observations up to time $n$,
i.e., $Y_n =\{y_1,\ldots ,y_n\}$.
The prediction problem is to obtain, $p(x_n|Y_{n-1})$,
 the conditional distribution of $x_n$ given $Y_{n-1}$, 
and the filter problem is to obtain, $p(x_n|Y_n)$, 
the conditional distribution of $x_n$ given $Y_n$.
For the linear state-space model with Gaussian mixture noise, 
it is known that these conditional distributions are also 
given as the mixture of Gaussian densities\cite{AS 1972},\cite{Kitagawa 1989},\cite{Kitagawa 1994},\cite{SA 1971}:
\begin{eqnarray}
 p(x_n|Y_{n-1}) &=& \sum_{i=1}^q \sum_{k=1}^{\ell_{n-1}}
   \alpha_i \gamma_{k,n-1} \varphi (x_n|x_{n|n-1}^{ik},V_{n|n-1}^{ik}) 
   = \sum_{j=1}^{m_n}
   \delta_{jn} \varphi (x_n|x_{n|n-1}^{j},V_{n|n-1}^{j})\nonumber \\
 p(x_n|Y_n) &=& \sum_{j=1}^r \sum_{k=1}^{m_n}
   \gamma_{jk,n} \varphi (x_n|x_{n|n}^{jk},V_{n|n}^{jk}) 
   = \sum_{i=1}^{\ell_n}
   \gamma_{in} \varphi (x_n|x_{n|n}^{i},V_{n|n}^{i})
\end{eqnarray}
where $m_n = q\times\ell_{n-1}$, $\delta_{jn}=\alpha_i \gamma_{k,n-1}$, 
$\ell_n = r\times m_n$ and $\gamma_{jk,n}=\beta_j\delta_{kn}\varphi(
y_n|x_{n|n-1}^{jk},V_{n|n-1}^{jk})$.

The Gaussian-sum filter is an algorithm to obtain 
these conditional densities recursively with time.
The advantage of the Gaussian-sum filter is that the parameters 
of the state distributions such as $\delta_{jn}$, $\gamma_{in}$, $x_{n|t}$, and $V_{n|t}$ 
are obtained by running the Kalman filters in parallel.
Therefore, the computation is easy and can yield accurate results.
However, there is a severe difficulties with this method.
Namely, the numbers of Gaussian components, $m_n$ and $\ell_n$, increase by $q\times r$ times 
at each time step of the filtering. 
Therefore, the number of Gaussian components would increase exponentially over time,
and for this filtering method to be practical, a computationally efficient method for 
the reduction of the number of Gaussian components is indispensable. 

In principle, reduction of the number of Gaussian components can be realized
by minimizing the Kullback-Leibler divergence of the full-order Gaussian mixture
distribution with respect to the reduced-order Gaussian mixture distribution.
However, as we discussed later in Section 2, two problems make this method impractical.
Therefore, as a practical measure, we usually reduce the number of Gaussian components successively.
In this paper, we refer to this method as the sequential reduction method
and consider criteria for selecting a pair of Gaussian components to be merged.

Kitagawa\cite{Kitagawa 1989}\cite{Kitagawa 1994} used a weighted Kullback-Leibler
divergence of two candidate Gaussian components.
Salmond\cite{Salmond 1990} proposed a mixture reduction algorithm in which the number of components is reduced by repeatedly choosing the two components that appear to be most similar to each other. 
Williams and Maybeck\cite{WM 2003} proposed a mixture reduction algorithm based on an integrated squared difference (ISD) similarity measure, which has the big advantage that the similarity between two arbitrary Gaussian mixtures can be expressed in closed form. 
Runnalls\cite{Runnalls 2007} proposed  a measure of similarity between two components based on the upper bound of the increase of Kullback-Leibler (KL) discrimination measure when a pair of two Gaussian components are merged. 
In this paper, we propose use of Pearson $\chi^2$-divergence of two Gaussian components 
for which we can derive a closed form expression for the criterion to select the pair of Gaussian components to be merged.

In section 2, we define the Gaussian mixture reduction problem and briefly 
show some reduction methods.
In section 3, a sequential reduction method based on Pearson $\chi^2$-divergence
will be introduced, in which the criteria for selecting a pair of indices to be
merged can be obtained in explicit analytical form.
In section 4, emperical studies on the sequential reduction of the number of
Gaussian components are shown, using one-dimensional and two-dimensional
Gaussian mixture distributions.
Section 5 deals with the application of the sequential Gaussian-mixture reduction 
method to the a Gaussianm-sum filtering and smoothing for linear state-space model
with Gaussian-mixture noise inputs.
We conclude in Section 6.
Details of the derivation of the Pearson $\chi^2$-divergence is shown in Appendix.

\section{Reduction of Gaussian Components}\index{reducion of gaussian components} 

\subsection{Reduction based on Kullback-Leibler Discrimination}\index{KL discrimination}  

The Kullback-Leibler divergence is the most frequently used to evaluate the
dissimilarity between true distribution and an approximated distribution,
which is defined by 
\begin{eqnarray}
I(g(x);f(x))  
  = \int \log \left\{ \frac{g(x)}{f(x)}\right\}g(x)dx
  = \int \log \left\{ {g(x)}\right\}g(x)dx - \int \log \left\{ {f(x)}\right\}g(x)dx, 
\end{eqnarray}
where in the context of the Gaussian mixture approximation, 
$g(x)$ is the full-order mixture model and $f(x)$ is the reduced order model ($\ell < m$):
\begin{eqnarray}
g(x) &=& \sum_{i=1}^m \alpha_i \varphi(x|\xi_i,V_i)\\
f_{\ell}(x) &=& \sum_{i=1}^{\ell} \beta_i \varphi(x|\mu_i,\Sigma_i).
\end{eqnarray}
Hereafter, for simplicity of the notation, the number of Gaussian components
is referred to as the order.

In principle, the best reduced order model can be obtained by 
minimizing the Kullback-Leibler divergence. 
However, there are two problems with this method. 
Firstly, except for simple densities such as Gaussian density, 
the KL-divergence does not have a closed expression. 
So we need to apply numerical integration to evaluate the KL-divergence. 
Secondly, to estimate the parameters of the best reduced order model, 
we need to apply numerical optimization in high dimensional parameter space.
Therefore, at least for recursive filtering in which this reduction process
is repeated as long as a new observation is obtained, this method is impractical.

\subsection{Sequential Reduction}
Therefore, we usually apply a sequential reduction method.
Assume that the full-order model and an approximated reduced order model
are respectively defined by 
\begin{eqnarray}
g(x) &=& \sum_{i=1}^m w_i \varphi(x|\xi_i,U_i) \nonumber \\
f_\ell(x) &=& \sum_{i=1}^\ell \alpha_i \varphi(x|\mu_i,\Sigma_i) .
\end{eqnarray}
In the sequential reduction method, to further reduce the number of components, 
we select a pair of two components, say $j$ and $k$, 
and pool these two densities. 
The reduced order model is defined by 
\begin{eqnarray}
h_{jk}(x) = \sum_{i\not\in \{j,k\}} \alpha_i \varphi(x|\mu_i,\Sigma_i) 
          + (\alpha_j + \alpha_k) \varphi(x|\zeta_{jk},V_{jk}) 
\end{eqnarray}
where $\varphi(x|\zeta_{jk},V_{jk})$ is the merged density 
whose parameters are usually determined 
so that the first two moments of the distributions are preserved:
\begin{eqnarray}
\xi_{jk} &=& (\alpha_j+\alpha_k)^{-1}\left(\alpha_j\mu_j+\alpha_k\mu_k\right) \label{eq_moment-preserving-merge}\\
V_{jk} &=& (\alpha_j+\alpha_k)^{-1}
        \left[\alpha_j\left\{\Sigma_j+(\mu_j-\xi_{jk})(\mu_j-\xi_{jk})^T\right\}
            + \alpha_k\left\{\Sigma_k+(\mu_j-\xi_{jk})(\mu_j-\xi_{jk})^T\right\}\right] .\nonumber
\end{eqnarray}

The indices of two pooled densities, $j$ and $k$, are selected so that 
a properly determined criterion is minimized.
By repeating this process, we can obtain a Gaussian mixture
approximation of $g(x)$ with a smaller number of Gaussian components.

For selecting a pair of two densities,
many ad hoc criteria have been proposed so far.
Kitagawa(1989,1994) used the weighted KL-divergence of Gaussian components
\begin{eqnarray}
D(k,j)=\alpha_k\alpha_j\left\{\Sigma_k^{-1}\Sigma_j +\Sigma_j^{-1}\Sigma_k
    + (\mu_k-\mu_j)^T(\Sigma_k^{-1}+\Sigma_j^{-1})(\mu_k-\mu_j)\right\}.
\end{eqnarray}
Salmond(1990) proposed the increase of within-component variance 
\begin{eqnarray}
    D_s^2(k,j) ={\rm tr}(\Sigma^{-1}\Delta W), \quad 
\Delta W(\varphi_k,\varphi_j)=\frac{\alpha_k\alpha_j}{\alpha_k+\alpha_j}(\mu_k-\mu_j)(\mu_k-\mu_j)^T.
\end{eqnarray}
Williams and Mayback (2003) used a squared difference of two densities
\begin{eqnarray}
    J(g,f) = \int (g(x)-f(x))^2 dx.
\end{eqnarray}
Runnalls(2006) used the upper bound of the increase of KL-divergence
by pooling two densities:
\begin{eqnarray}
B(k,j) = \frac{1}{2}\left\{(\alpha_k+\alpha_j)
    \log \det (V_{kj}) - \alpha_k \log\det (\Sigma_k) -\alpha_j\log\det (\Sigma_j)    \right\}
\end{eqnarray}
and it is reported that this criterion mitigated some anomalous behavior in
certain circumstances of the ones by Williums and Mayback\cite{WM 2003} 
and Salmond\cite{Salmond 1990}, 
and provide us with a reasonable reduction result\cite{Runnalls 2007}.


\section{Reduction Criterion based on Pearson $\chi^2$-Divergence}\index{pearson divergence}

\subsection{Pearson $\chi^2$-Divergence of Two gaussian Mixture Models}
In this paper, we consider the use of Pearson $\chi^2$-divergence:
\begin{eqnarray}
D_{\chi^2}(q;p) = \int \left(\frac{q(x)}{p(x)}-1\right)^2 p(x)dx 
    =\int \frac{q(x)^2}{p(x)}dx -1.
\end{eqnarray}
Assume that $q(x)$ is a mixture of two Gaussian densities
\begin{eqnarray}
q(x) = \alpha_j \varphi (x|\mu_j,\Sigma_j) + \alpha_k \varphi (x|\mu_k,\Sigma_k), \quad \alpha_j + \alpha_k = 1 
\end{eqnarray}
and $p(x)$ is a pooled Gaussian density, $p_{jk}(x) = \varphi (x|\zeta_{jk},W_{jk})$, obtained by
the moment preserving merge 
where $\zeta_{jk}$ and $V_{jk}$ are given in (\ref{eq_moment-preserving-merge}).
Then the Pearson $\chi^2$-divergence $D_{\chi^2}(j,k)$ of the mixture of two Gaussian
densities with respect to the merged density is obtained by
\begin{eqnarray}
D_{\chi^2}(j,k) &=& \int \frac{q(x)^2}{p_{jk}(x)} dx -1 \nonumber \\
 &=& \alpha_j^2 \int \frac{f_j(x)^2}{p_{jk}(x)}dx  
 + 2\alpha_j\alpha_k \int \frac{f_j(x)f_k(x)}{p_{jk}(x)} dx  
 + \alpha_k^2 \int \frac{f_k(x)^2}{p_{jk}(x)} dx -1 . \label{Perason_divergence}
\end{eqnarray}

Here, since the densities $f_j(x)$, $f_k(x)$ and $p_{jk}$ are respectively defied by
\begin{eqnarray}
f_j(x) &=& (2\pi )^{-\frac{k}{2}}\left|\Sigma_j\right|^{-\frac{1}{2}}
  \exp\left\{ -\frac{1}{2}(x-\mu_j)^T\Sigma_j^{-1} (x-\mu_j)\right\} \nonumber\\
f_k(x) &=& (2\pi )^{-\frac{k}{2}}\left|\Sigma_k\right|^{-\frac{1}{2}}
   \exp\left\{ -\frac{1}{2}(x-\mu_k)^T\Sigma_k^{-1} (x-\mu_k)\right\} \\
p_{jk}(x) &=& (2\pi )^{-\frac{k}{2}}\left|V_{jk}\right|^{-\frac{1}{2}}
   \exp\left\{ -\frac{1}{2}(x-\zeta_{jk})^TV_{jk}^{-1}(x-\zeta_{jk})\right\}, \nonumber 
\end{eqnarray}
the integrand of the second term of the right hand side 
of the equation (\ref{Perason_divergence}) is given by
\begin{eqnarray}
\lefteqn{ \frac{f_j(x)f_k(x)}{p_{jk}(x)}
 = (2\pi )^{-\frac{k}{2}}\left|\Sigma_j\right|^{-\frac{1}{2}}\left|\Sigma_k\right|^{-\frac{1}{2}}\left| V_{jk}\right|^{\frac{1}{2}} } \nonumber \\
 && \times \exp\left\{ -\frac{1}{2}(x-\mu_j)^T\Sigma_j^{-1} (x-\mu_j) 
              -\frac{1}{2}(x-\mu_k)^T\Sigma_k^{-1} (x-\mu_k)
              +\frac{1}{2}(x-\xi_{jk})^T V_{jk}^{-1} (x-\xi_{jk})
\right\} \nonumber \\
&=& (2\pi )^{-\frac{k}{2}}\left|\Sigma_j\right|^{-\frac{1}{2}}\left|\Sigma_k\right|^{-\frac{1}{2}}\left| V_{jk}\right|^{\frac{1}{2}} 
\exp\left\{-\frac{1}{2}(\mu_j - \mu_k)^T (\Sigma_j + \Sigma_k)^{-1}(\mu_j - \mu_k)\right\}\nonumber \\
 && \times\exp\left\{ -\frac{1}{2}(\zeta_{jk} -\eta_{jk})^T (V_{jk}-\Sigma_{jk} )^{-1}(\zeta_{jk} -\eta_{jk} )\right\} 
          \exp\left\{ -\frac{1}{2}(x-\eta_{jk} )^T W_{jk} (x-\eta_{jk}) \right\} \label{eq-integrand_f1(x)f2(x)}
\end{eqnarray}
where $\Sigma_{jk} = (\Sigma_j^{-1}+\Sigma_k^{-1})^{-1}$, $W_{jk} = \Sigma_j^{-1}+\Sigma_k^{-1}-V_{jk}^{-1}$ and $\eta_{jk} = (\Sigma_j^{-1}+\Sigma_k^{-1}-V_{jk}^{-1})^{-1}((\Sigma_j^{-1}+\Sigma_k^{-1})\zeta_{jk}-V_{jk}^{-1}\xi_{jk} )$.
The details of the derivation of the last equality of (\ref{eq-integrand_f1(x)f2(x)}) is given in the appendix.

Then, by integrating over the whole domain of the distribution, we obtain
\begin{eqnarray}
  \int \frac{f_j(x)f_k(x)}{p_{jk}(x)} dx 
&=& \left|\Sigma_j\right|^{-\frac{1}{2}}
  \left|\Sigma_k\right|^{-\frac{1}{2}}
  \left|V_{jk}\right|^{\frac{1}{2}}  \left| W_{jk} \right|^{-\frac{1}{2}} 
  \exp \left\{ -\frac{1}{2}
   (\zeta_{jk} -\eta_{jk})^T (V_{jk}-\Sigma_{jk} )^{-1} (\zeta_{jk} - \eta_{jk}) \right\}  \nonumber\\
&& \times \exp \left\{ -\frac{1}{2}
   (\mu_j-\mu_k)^T (\Sigma_j+\Sigma_k)^{-1} (\mu_j - \mu_k) \right\} .\label{eq_integral_of fjfk_over_p}
\end{eqnarray}
The expression for the first and the third term of (\ref{Perason_divergence}) is obtained by 
putting by $f_k(x) = f_j(x)$;
namely, $\mu_k=\mu_j$ and $\Sigma_k=\Sigma_j$.
\begin{eqnarray}
  \int \frac{f_j(x)^2}{p_{jk}(x)} dx 
= \left|\Sigma_j\right|^{-1}  \left|V_{jk} \right|^{\frac{1}{2}} 
  \left|\bar{W}_j\right|^{-\frac{1}{2}} 
\exp \left\{ -\frac{1}{2}(\mu_j-\eta_{jk})^T\bar{W}_j^{-1}(\mu_j-\eta_{jk})\right\} 
\end{eqnarray}
where $\bar{W}_j = 2\Sigma_j^{-1}-V_{jk}^{-1}$, $\eta_j = (2\Sigma_j^{-1}-V_{jk}^{-1})^{-1}(2\Sigma_j^{-1}\mu_j -V_{jk}^{-1}\xi_{jk} )$.

\subsection{Proposed Reduction Criterion}

Therefore the Pearson $\chi^2$-divergence for the Gaussian mixture reduction is obtained by
\begin{eqnarray}
D_{\chi^2}(j,k) 
 &=& \alpha_j^2 \left|\Sigma_j\right|^{-1}  \left|V_{jk} \right|^{\frac{1}{2}} \left|\bar{W}_j\right|^{-\frac{1}{2}} 
\exp \left\{ \frac{1}{2}(\mu_j-\xi_{jk})^T(V_{jk}-\frac{1}{2}\Sigma_j)^{-1}(\mu_j-\xi_{jk})\right\} 
 \nonumber \\
&+& \alpha_k^2 \left|\Sigma_k\right|^{-1}  \left|V_{jk} \right|^{\frac{1}{2}} \left|\bar{W}_k\right|^{-\frac{1}{2}} 
\exp \left\{ -\frac{1}{2}(\mu_k-\xi_{jk})^T(V_{jk}-\frac{1}{2}\Sigma_k)^{-1}(\mu_k-\xi_{jk})\right\} 
 \nonumber \\
&+& 2\alpha_i\alpha_j 
  \left|\Sigma_j\right|^{-\frac{1}{2}} \left|\Sigma_k\right|^{-\frac{1}{2}}
   \left|V_{jk} \right|^{\frac{1}{2}} \left| W_{jk} \right|^{-\frac{1}{2}}
   \exp\left\{ -\frac{1}{2} (\zeta_{jk} -\xi_{jk})^T (V_{jk}-\Sigma_{jk})^{-1} (\zeta_{jk} - \xi_{jk}) \right\} \nonumber\\
&& \hspace{20mm}\times \exp \left\{ -\frac{1}{2}
   (\mu_j-\mu_k)^T (\Sigma_j+\Sigma_k)^{-1} (\mu_j - \mu_k) \right\}
   -1
\end{eqnarray}
In the sequential reduction based on this criterion, $D_{\chi^2}(j,k)$ are
evaluated for $j=1,...,\ell -1$ and $k=2,...,\ell $ and
find the pair $(j^*,k^*)$ that satisfies
\begin{eqnarray}
  D_{\chi^2}(j^*,k^*) = \min_{j,k} D_{\chi^2}(j,k).
\end{eqnarray}
Then the two Gaussian components $\varphi (x|\mu_j^*,\Sigma_j^*)$
and $\varphi (x|\mu_k^*,\Sigma_k^*)$ are merged and we obtain
the Gaussian mixture model with $\ell -1$ components.
Repeating this process, it is possible to obtain Gaussian mixture
distribution with a specific order.

The problem with this Pearson $\chi^2$-divergence is that
$q(x)/p(x)$ may become unbounded. 
Therefore, in using this as the  criterion for selecting the
pair for merging, we need a safe-guard in computation.
Namely, we exclude the pair $j$ and $k$ from the merging candidate.

\section{Empirical Study: Comparison of Reduction Methods}
Many criteria have been proposed for selecting a pair of
Gaussian components in sequential reduction of Gaussian components.
In this section we compare the following criteria:  
\begin{enumerate}
\item Weighted KL-divergence of Gaussian components, Kitagawa (1989, 1994):  
\begin{eqnarray}
D(j,k) = \alpha_j\alpha_k \left\{ \Sigma_j^{-1}\Sigma_k + \Sigma_j^{-1}\Sigma_k
 + (\mu_j -\mu_k)^T(\Sigma_j^{-1}+\Sigma_k^{-1})(\mu_j-\mu_k)\right\}
\end{eqnarray}
\item Upper bound of the increase of KL-divergence, Runalls (2006): 
\begin{eqnarray}
B(j,k) = \frac{1}{2}\left\{ (\alpha_j+\alpha_k)\log \det (V_{jk})
  -\alpha_j \log \det (\Sigma_j) -\alpha_k \log \det (\Sigma_k)  \right\}
\end{eqnarray}
\item $\chi^2$-divergence proposed in this paper: $D_{\chi^2}(j,k)$
\end{enumerate}
Beside these ad hoc criteria, we also considered the following two
reduction methods based on the Kullback-Leibler divergence.
\begin{enumerate}
\setcounter{enumi}{3}
\item The sequential reduction based on the Kullback-Leibler
divergence of the pooled model obtained by numerical integration:
\begin{eqnarray}
I(g;f_{jk}) = \int \log g(x)g(x)dx - \int \log f_{jk}(x)g(x)dx.
\end{eqnarray}
\item The global Kullback-Leibler divergence minimization method.
Note that this method requires both numerical integration and 
numerical optimization:
\begin{eqnarray}
I(g;\hat{f}_{jk}) = \int \log g(x)g(x)dx - \int \log \hat{f}_{jk}(x)g(x)dx,
\end{eqnarray}
where the parameters of $f_{jk}(x)$ are estimated by minimizing $I(g;f_{jk})$.
Therefore, this method is very computationally costly
and is feasible only for very low dimensional distributions.
\end{enumerate}

\subsection{One-dimensional Distributions}

\begin{figure}[tbp]
\begin{center}
\includegraphics[width=100mm,angle=0,clip=]{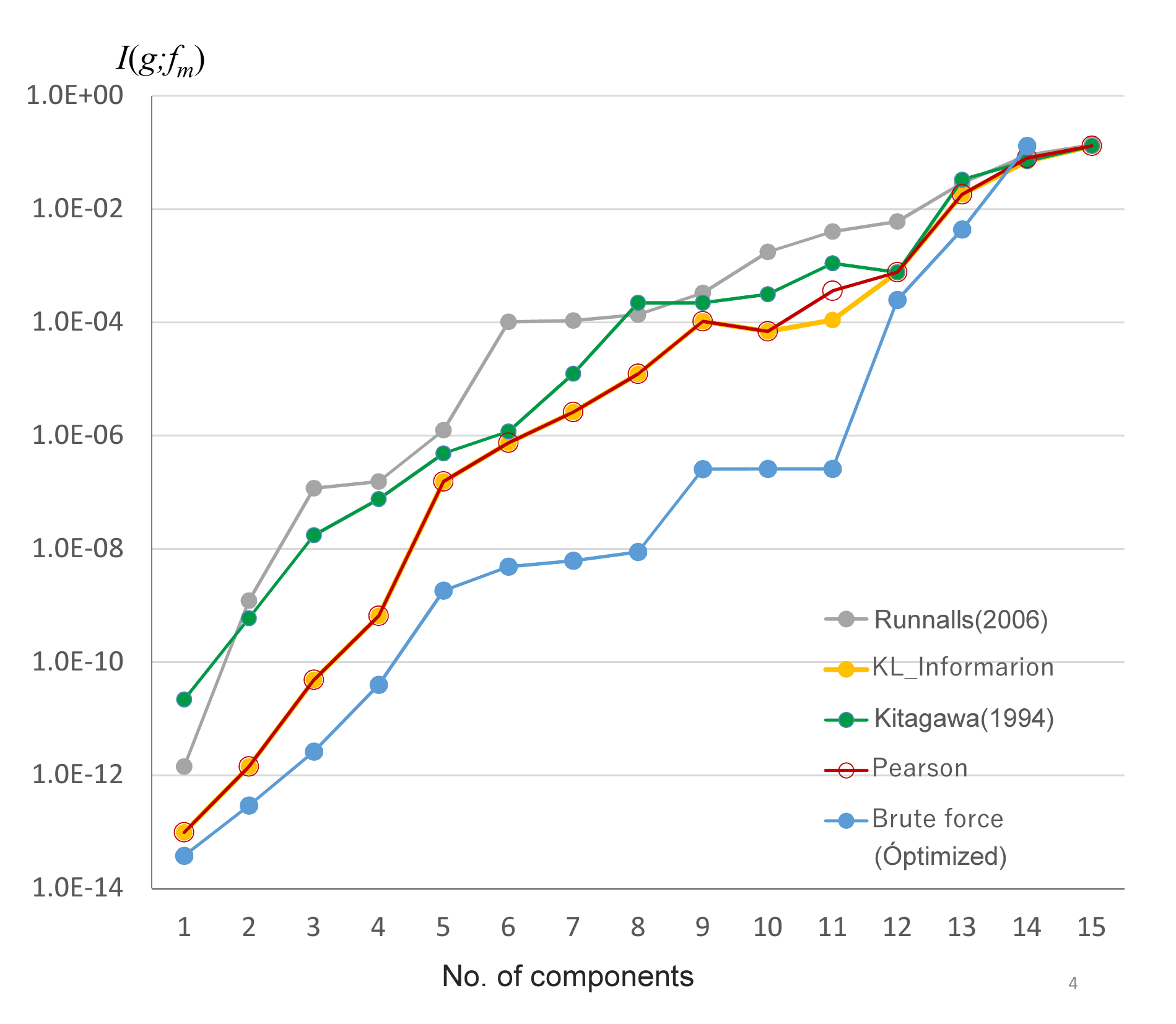}
\end{center}
\caption{Change in KL-divergence of true, sequentially reduced and optimal reduced order models. }\label{Fig_1D_reduction_comparison}
\vspace{5mm}
\end{figure}

Table \ref{One-dim_true_model} shows the assumed full-order Gaussian mixture model
with 16 Gaussian components.
\begin{table}[tbp]
\caption{Assumed one-dimensional Gaussian-mixture distribution
with 16 components.}\label{One-dim_true_model}
\begin{center}
\begin{tabular}{c|lcc} 
$i$ & {}\hspace{3mm}$\alpha_i$ & $\mu_i$ & $\Sigma_i$ \\\hline
1 &	0.30  & 0.0	 &0.5  \\
2 &	0.15  &	5.0	 &1.0   \\
3 &	0.15  &	-4.0 &	1.0  \\
4 &	0.05  &	0.2	 & 9.0  \\
5 &	0.05  &	-1.5 &	2.0  \\
6 &	0.0686  &  1.03982 &	4.39842  \\
7 &	0.03472 & -1.55209 &	3.78821  \\
8 &	0.07578	& -1.35090 &	2.78963  \\
9 &	0.00101	& -0.25711 &	1.18460  \\
10 &0.00011	& 2.00426 &	1.14186  \\
11 &0.01699	& 1.44357 &	1.00000  \\
12 &0.00003	&-2.15010 &	1.02979  \\
13 &0.05787	&-0.58808 &	1.21395  \\
14 &0.00039	& 1.57966 &	1.35196  \\
15 &0.02193	& 1.87170 &	1.12458  \\
16 &0.02257	& 0.55285 &	1.05299  \\
\hline
\end{tabular}
\end{center}
\end{table}
Table \ref{Tab-KL-divergence} and Figure \ref{Fig_1D_reduction_comparison} 
show the increase of KL-divergence 
when the reduced order models are obtained by five methods. 
In the figure, grey line shows the results by Runnalls, green one by Kitagawa, 
blue one by Pearson $\chi^2$-divergence, yellow one sequential reduction 
by Kullback-Leibler divergence, and red one by global optimization
of Kullback-Leibler divergence.
It can be seen that the sequential reduction based on Pearson $\chi^2$-divergence 
yields almost the same performance as the sequential reduction by
Kullback-Leibler divergence.

The accuracy of the sequential reduction methods are worth 
by one or two digit than the optimal model. 
However, the figure also indicates that by using a larger order $m$, 
we can attain a similar accuracy as the optimal model.

\begin{table}[tbp]
\caption{Change of KL-divergence of true with respect to the  
reduced order models by various reduction methods.}\label{Tab-KL-divergence}
\begin{center}
\begin{tabular}{r|ccccc} \hline
$m$ & Runnalls & Kitagawa & Pearson & KL-div. & Optimal \\ \hline
 15& 1.43$\times 10^{-12}$& 2.18$\times 10^{-11}$& 9.80$\times 10^{-14}$& 3.00$\times 10^{-13}$&  3.80$\times 10^{-14}$\\
 14& 1.21$\times 10^{-09}$& 5.98$\times 10^{-10}$& 1.43$\times 10^{-12}$& 4.63$\times 10^{-12}$&2.94$\times 10^{-13}$\\
 13& 1.17$\times 10^{-07}$& 1.74$\times 10^{-08}$& 4.85$\times 10^{-11}$& 3.40$\times 10^{-11}$&2.63$\times 10^{-12}$\\
 12& 1.54$\times 10^{-07}$& 7.55$\times 10^{-08}$& 6.53$\times 10^{-10}$& 3.24$\times 10^{-11}$& 3.96$\times 10^{-11}$\\
 11& 1.24$\times 10^{-06}$& 4.85$\times 10^{-07}$& 1.54$\times 10^{-07}$& 1.78$\times 10^{-09}$&1.82$\times 10^{-09}$\\
 10&  0.00010181& 1.18$\times 10^{-06}$& 7.45$\times 10^{-07}$& 3.70$\times 10^{-09}$&4.82$\times 10^{-09}$\\
  9&  0.00010793& 1.24$\times 10^{-05}$& 2.60$\times 10^{-06}$& 1.08$\times 10^{-08}$&6.15$\times 10^{-09}$\\
  8&  0.00013676&  0.00022274& 1.23$\times 10^{-05}$& 1.67$\times 10^{-08}$&8.84$\times 10^{-09}$\\
  7&  0.00033167&  0.00022197&0.0001042& 2.88$\times 10^{-07}$&  2.55$\times 10^{-07}$\\
  6&  0.00175442&  0.00031239& 6.90$\times 10^{-05}$& 2.66$\times 10^{-07}$&  2.57$\times 10^{-07}$\\
  5&   0.0040189&  0.00110572&  0.00035793&   1.61$\times 10^{-06}$&  2.57$\times 10^{-07}$\\
  4&   0.0060584&  0.00076506&  0.00076506&  0.00024942&  0.00024942\\
  3&  0.02886692&   0.0331135&  0.01810894&  0.01650889&  0.00435254\\
  2&  0.08941172&  0.07007295&  0.07938004&  0.06884198&  0.06884198\\
  1&  0.13589858&   0.1304686&   0.1304686&   0.1304686&   0.1304686\\
\hline
\end{tabular}
\end{center}
\end{table}

\begin{figure}[tbp]
\begin{center}
\includegraphics[width=100mm,angle=0,clip=]{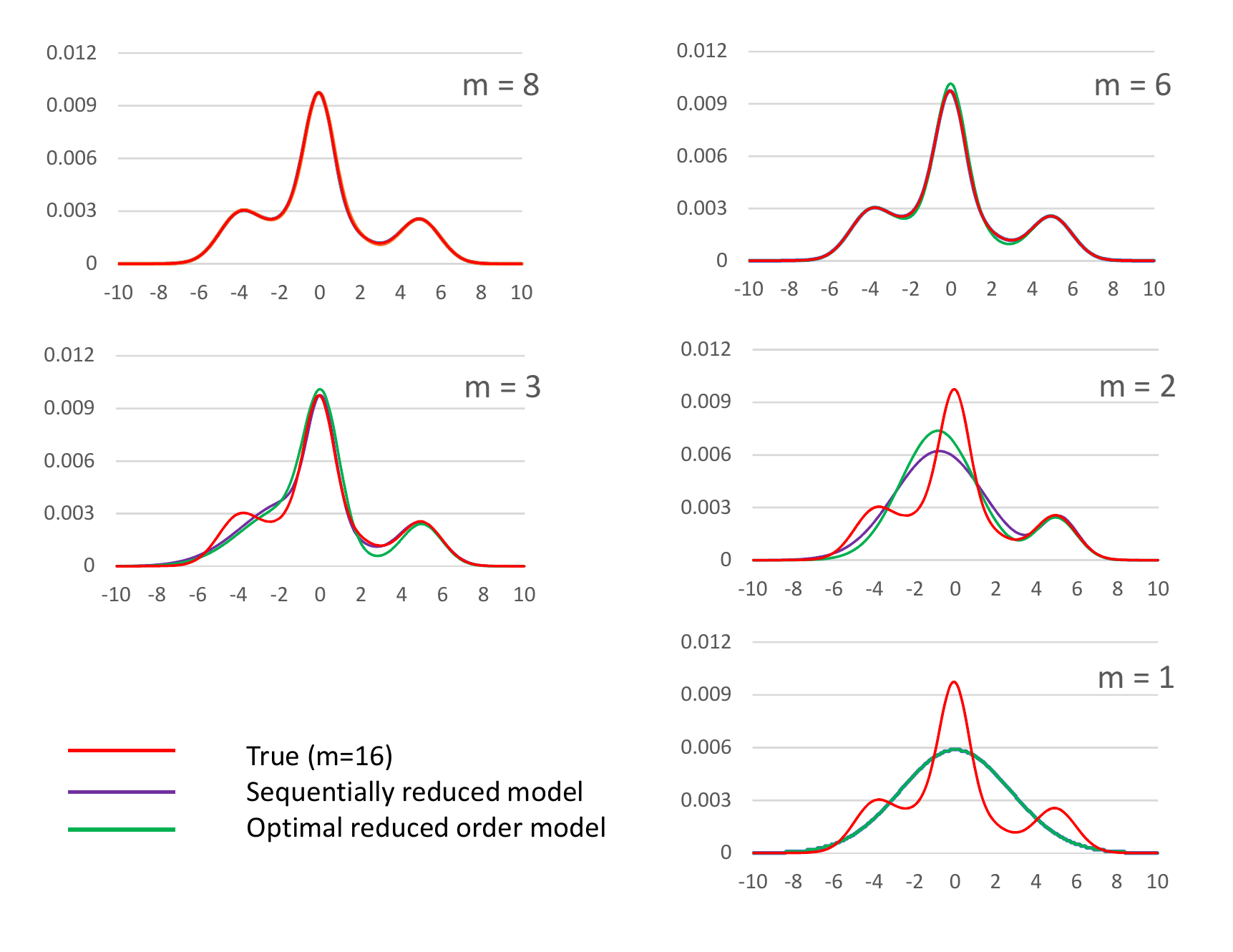}
\end{center}
\caption{The comparison of the densities obtained by the sequential 
reduction and the global optimization method. }\label{Fig_sequential_reduction}
\vspace{5mm}
\end{figure}

Figure \ref{Fig_sequential_reduction} shows the comparison of the densities obtained by the sequential 
reduction and the global optimization method.
In these plots, the red curve shows the true full order density, the green one the optimal reduced order model obtained by minimizing the KL-divergence, and the purple one obtained by the  sequential reduction based on the Pearson $\chi^2$-divergence.
It can be seen that for $m\geq 8$, the green curve and purple curve are visually indistinguishable. But for $m$=2 and 3, they are considerably different.

\newpage
\subsection{Two-dimensional Distributions}

In this example, the true 2-dimensional density is expressed by
10 Gaussian distributions shown in Table \ref{Two-dim_true_model}.
Table \ref{Tab-KL-divergence_2} and Figure \ref{Fig_Gaussian-sum-with-m=1} show
the Kullback-Leibler divergence of the true mixture model with respect to
the reduced order model obtained by 5 methods.
It can be seen that, except for $\ell$=2 and 3, the results 
by the Pearson $\chi^2$-divergence is almost indistinguishable
with the method based on Kullback-Leibler divergence.

\begin{table}[tbp]
\caption{Assumed twe-dimensional Gaussian-mixture distribution
with 16 terms.}\label{Two-dim_true_model}
\begin{center}
\begin{tabular}{c|rrrrrr} 
$i$ & $\alpha_i$ & $\mu_i(1)$ & $\mu_i(2)$ & $\Sigma_i(1,1)$ &$\Sigma_i(2,2)$ & $\Sigma_i(2,1)$ \\\hline
 1&   0.30  &     0  &     0    &   1   &    1  &     0  \\                
 2&   0.20  &     2  &     0    &   4   &    2  &     0  \\         
3&    0.16  &     3  &     3    &   2   &    2  &  -0.5  \\         
4&    0.11  &    -4  &    -4    &   4   &    4  &     2  \\      
5&    0.08  &    -1  &     1    &   9   &    9  &   4.0  \\    
6&    0.06  &     2  &    -4    &   4   &    9  &     2 \\      
7&    0.04  &     0  &     2    &   4   &    1  &  -0.5  \\    
8&    0.03  &    -2  &     4    &   9   &    9  &     0  \\ 
9&    0.01  &    -2  &     0    &   2   &    1  &     0  \\
10&   0.01  &     1  &    -2    &   1   &    1  &     0  \\      
\hline
\end{tabular}
\end{center}
\end{table}

\begin{figure}[tbp]
\begin{center}
\includegraphics[width=130mm,angle=0,clip=]{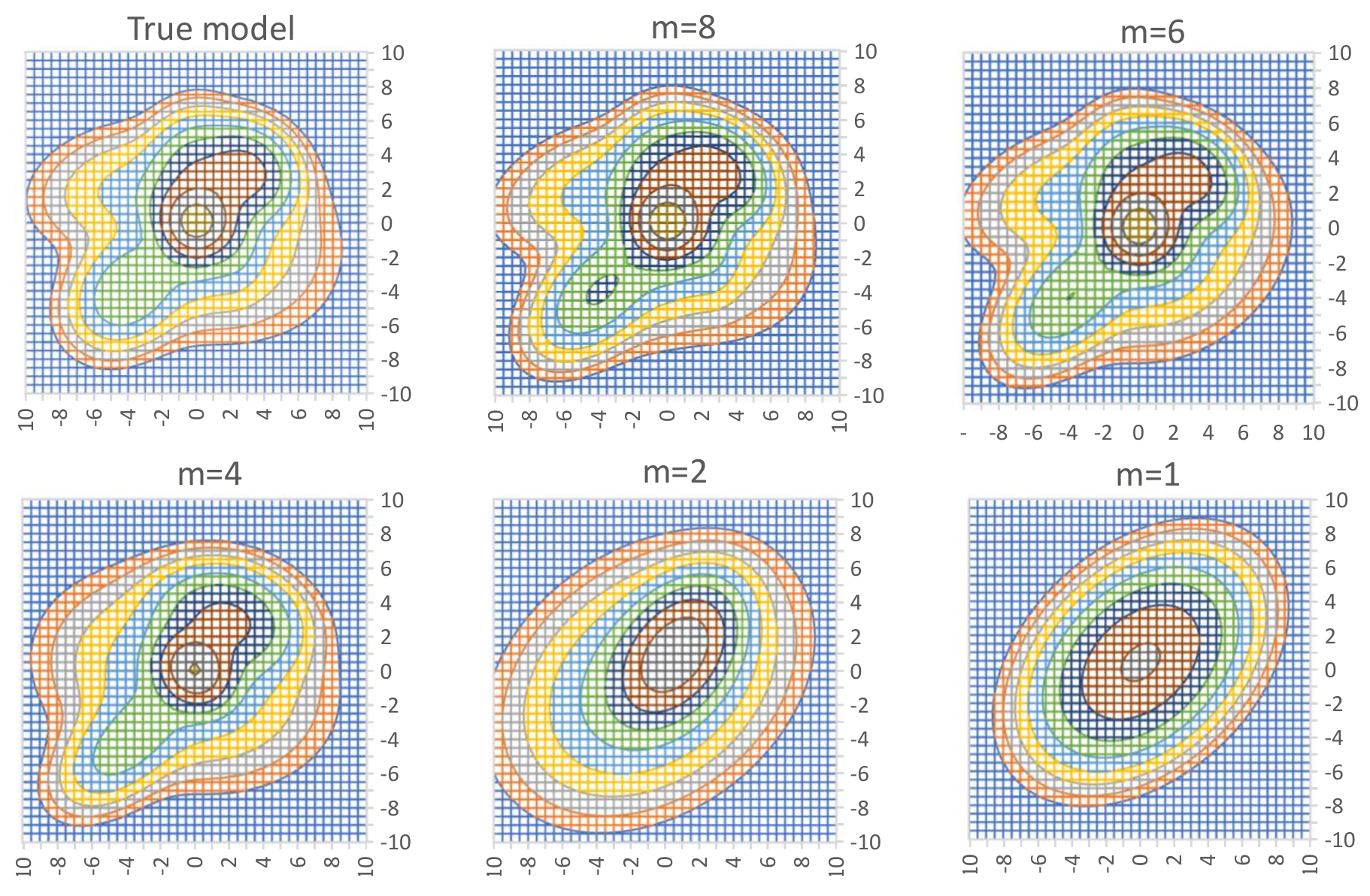}
\end{center}
\caption{Contour of 2D densities obtained from the full-order Gaussian-mixture
and reduced order Gaussian-mixture models. }\label{Fig_Contour_2D}
\vspace{5mm}
\end{figure}

\begin{figure}[ht]
\begin{center}
\includegraphics[width=140mm,angle=0,clip=]{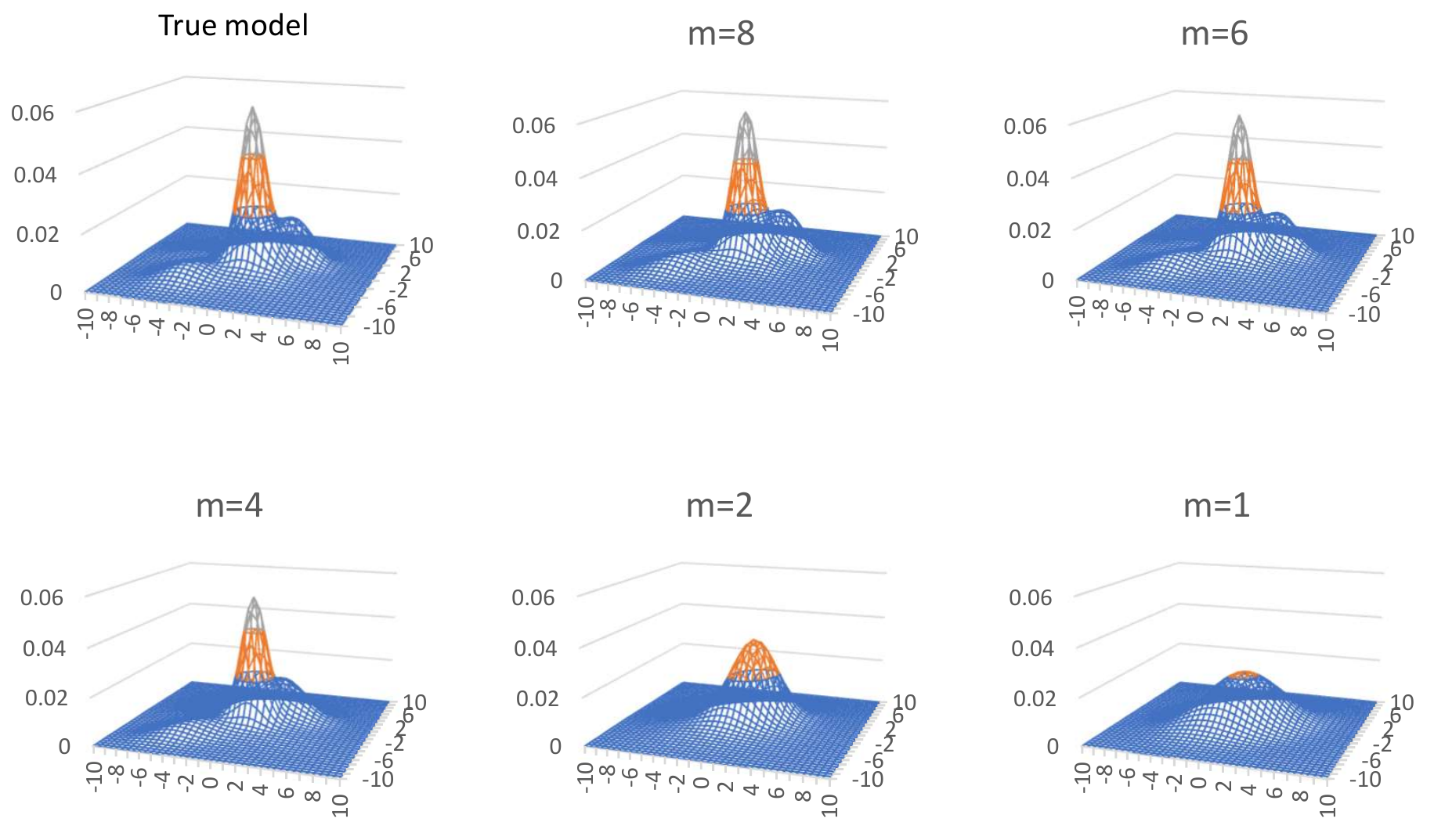}
\end{center}
\caption{Bird's-eye-views of 2D densities obtained from the full-order Gaussian-mixture and reduced order Gaussian-mixture models. }\label{Fig_2D_bird-eye}
\vspace{5mm}
\end{figure}

Figures \ref{Fig_Contour_2D} and \ref{Fig_2D_bird-eye} show the contour and the bird's-eye views of the
reduced order Gaussian mixture models obtained by the Pearson
$\chi^2$-divergence.

\begin{table}[tbp]
\caption{Change of KL-divergence of true model with respect to the  
reduced order models by various reduction methods:
Two dimensional case.}\label{Tab-KL-divergence_2}
\begin{center}
\begin{tabular}{r|ccccc} \hline
$m$ & Runnalls & Kitagawa & Pearson & KL-div. & Optimal \\ \hline
  9&  0.000220 & 0.000143 & 0.000163& 0.000143 & 0.000022 \\
  8&  0.000656 & 0.000849 & 0.000300& 0.000300 & 0.000093 \\
  7&  0.002367 & 0.001812 & 0.001051& 0.001051 & 0.000258 \\
  6&  0.004783 & 0.003920 & 0.002010& 0.002010 & 0.000496 \\
  5&  0.006878 & 0.023910 & 0.005754& 0.005754 & 0.002862 \\
  4&  0.029877 & 0.029670 & 0.014775& 0.014775 & 0.004916 \\
  3&  0.056387 & 0.034783 & 0.079955& 0.039786 & 0.029775 \\
  2&  0.099586 & 0.099586 & 0.122572& 0.091505 & 0.084608 \\
  1&  0.180119 & 0.180119 & 0.180119& 0.180119 & 0.180119\\
\hline
\end{tabular}
\end{center}
\end{table}

\begin{figure}[tbp]
\begin{center}
\includegraphics[width=95mm,angle=0,clip=]{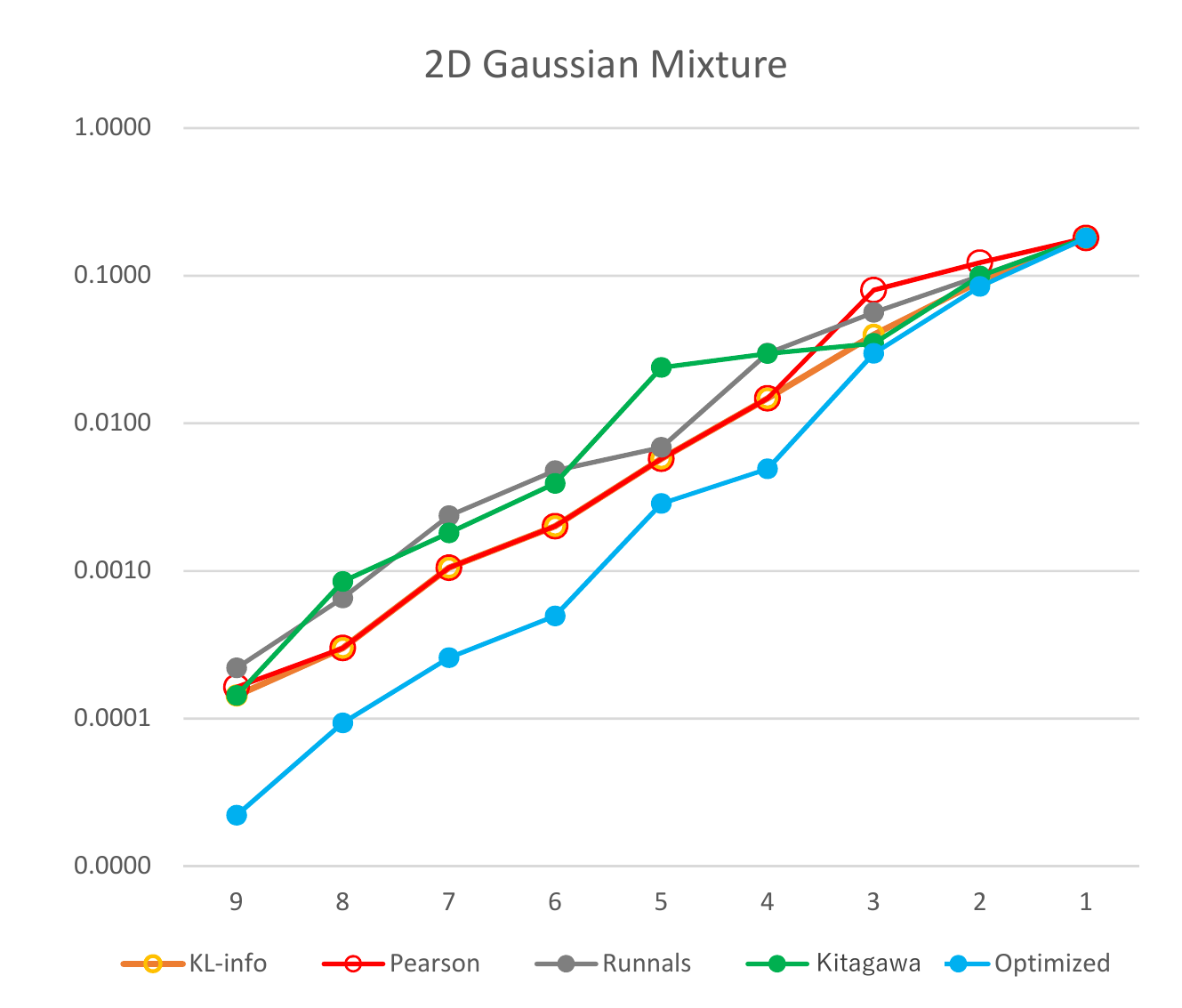}
\end{center}
\caption{Change in KL-divergence of true and optimal reduced order model. }
\label{Fig_2D_reduction_comparison}
\vspace{5mm}
\end{figure}

Summarizing the two examples, there are three types of reduction methods, 
namely the sequential reduction by ad-hoc criterion, 
Sequential reduction by KL-divergence and global KL-divergence minimization.
Obviously the accuracy increases in this order, 
but computational cost increases.
So the suggestion is to estimate a mixture model with a slightly larger 
number of components by the sequential reduction method.

\newpage

\section{Non-Gaussian Smoothing}
\begin{figure}[tbp]
\begin{center}
\includegraphics[width=150mm,angle=0,clip=]{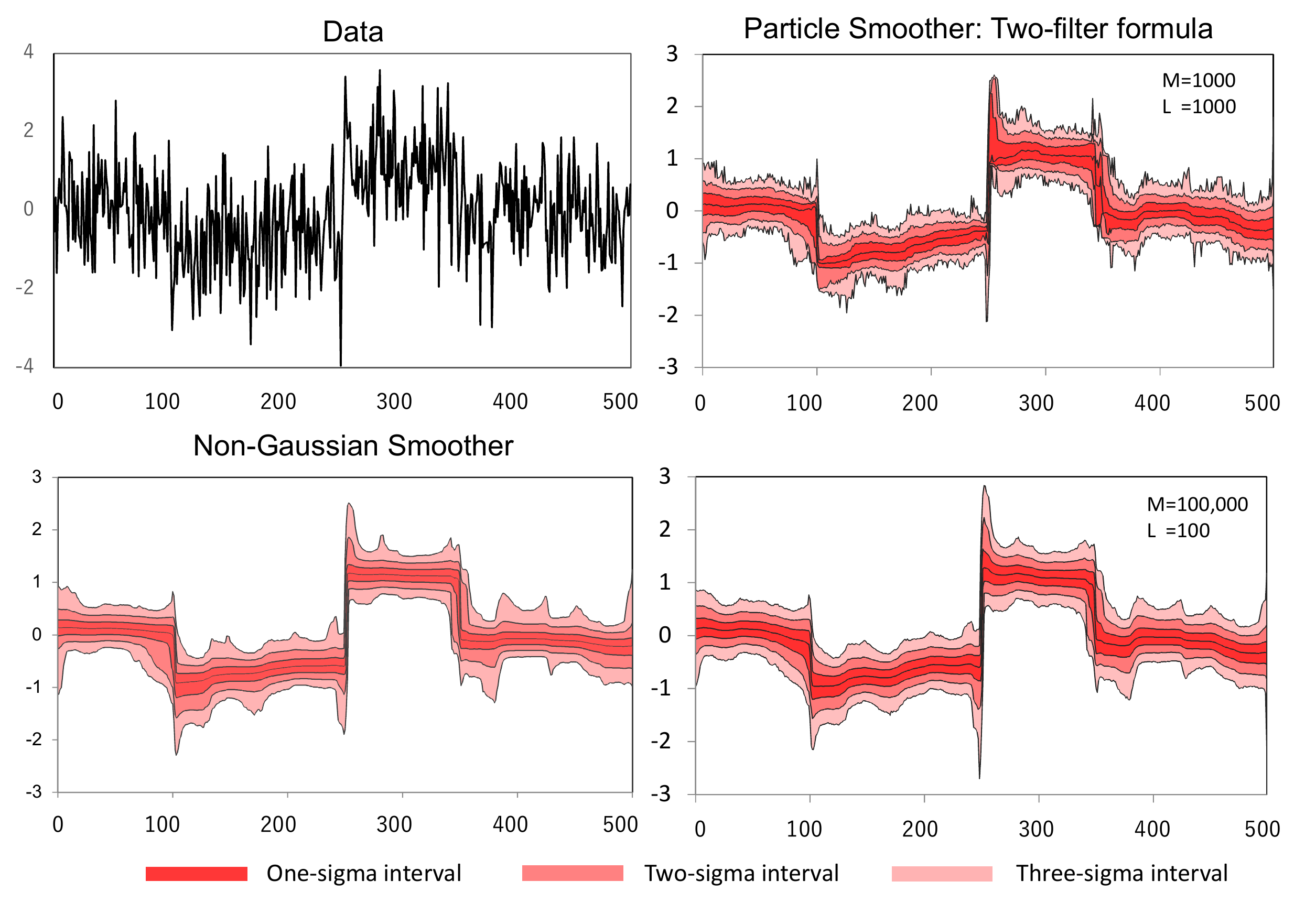}
\end{center}
\caption{Test data and the estimated trends obtained by the non-Gaussian smoother
and the particle filter with $m$=1000 and 100,000. }\label{Fig_particle_smoothers}
\vspace{5mm}
\end{figure}

We consider the application of Gaussian-sum filter and smoother 
to the detection of the level shift in the time series.
The top-left plot of Figure \ref{Fig_particle_smoothers} shows the example data analyzed in Kitagawa\cite{Kitagawa 1987}. 
For estimation of the trend of the series, we consider a simple state-space model. 
\begin{eqnarray}
x_n &=& x_{n-1} + v_n \nonumber\\
y_n &=& x_n + w_n.
\end{eqnarray}
Here we assume that the observation noise is Gaussian 
but the system noise is a mixture of two Gaussian distributions:
\begin{eqnarray}
v_n &\sim& \alpha N(0,\tau^2) + (1-\alpha )N(0,\xi^2) \nonumber \\
w_n &\sim& N(0,\sigma^2),
\end{eqnarray}
where $\sigma^2 = 1.027$, $\tau^2 = 0.000254$, $\xi^2 =1.189$ and $\alpha = 0.989$.

Figure \ref{Fig_particle_smoothers} show the estimates of the trend by the Non-Gaussian smoother \cite{Kitagawa 1987}
and the particle smoother \cite{Kitagawa 1996}.
Table \ref{Tab_Gaussian-sum-filter} shows the log-likelihoods and the cpu-times for various number of the maximum number of Gaussian components approximating the state 
densities.
At least in this case $M=8$ or 16 looks sufficient. 
The cpu-time is less than 1 second for filtering.

\begin{table}[tbp]
\caption{Gaussian-sum filters and smoothers for various number of Gaussian components.}
\label{Tab_Gaussian-sum-filter}
\begin{center}
\begin{tabular}{c|c|r|r} \hline
    &        & \multicolumn{2}{|c}{cpu time (in second)} \\
$m$ & log-lk & Filtering & Smoothing \\ \hline
  1 & -741.930 & 0.00 & 0.08\\ 
  2 & -741.047 & 0.02 & 0.23\\
  4 & -740.816 & 0.02 & 0.94\\
  8 & -740.748 & 0.05 & 3.70\\
 16 & -740.702 & 0.27 & 14.85\\
 32 & -740.704 & 1.86 & 59.53\\
 64 & -740.704 & 14.26 & 243.47\\
128 & -740.704 & 112.51& 1018.20\\
\hline
\end{tabular}
\end{center}
\end{table}

\begin{figure}[tbp]
\begin{center}
\includegraphics[width=150mm,angle=0,clip=]{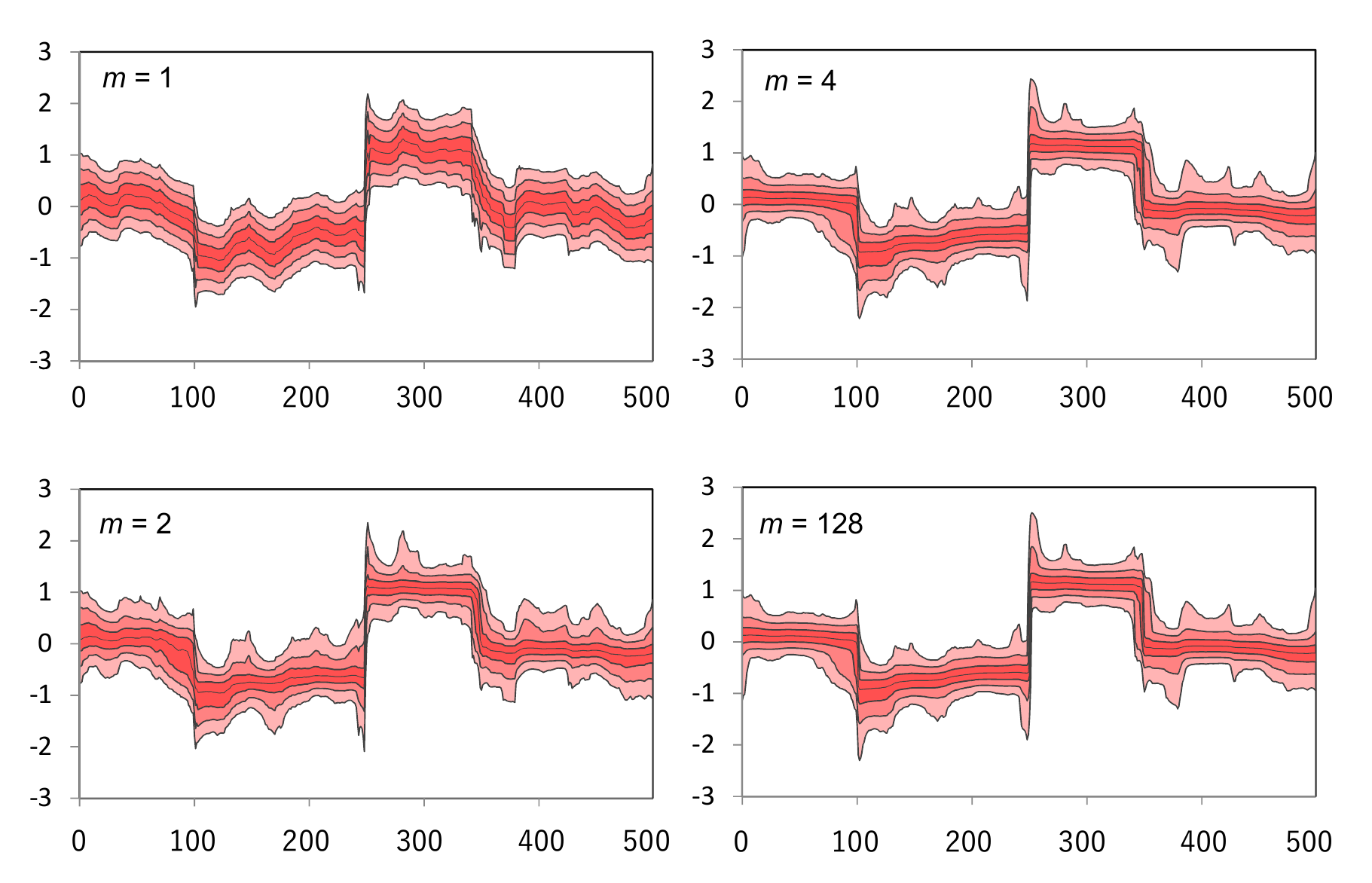}
\end{center}
\caption{Estimated trends by the Gaussian-sum smoother with number of Gaussian
components, $m$=1,2,4 and 128. }
\label{Fig_comparison-reduced-order-models}
\vspace{5mm}
\end{figure}

Figure \ref{Fig_comparison-reduced-order-models} shows the smoothed distribution 
of the trend obtained by the Gaussian-sum smoother for the number of components $m$=1,
2, 4 and 128.
The top-left plot shows the case $m=1$, bottom-left shows case $m=2$, 
top-right $m=4$ and bottom-right $m=128$. 
At least visually the results by $m=4$ and 128 are almost indistinguishable.
This indicates that the Gaussian-sum filter is very efficints for linear state
space model with Gaussian-mixture noise inputs in the sense that it can
provide a very accurate approximation to the posterior distribution of 
the state. 

\begin{figure}[tbp]
\begin{center}
\includegraphics[width=150mm,angle=0,clip=]{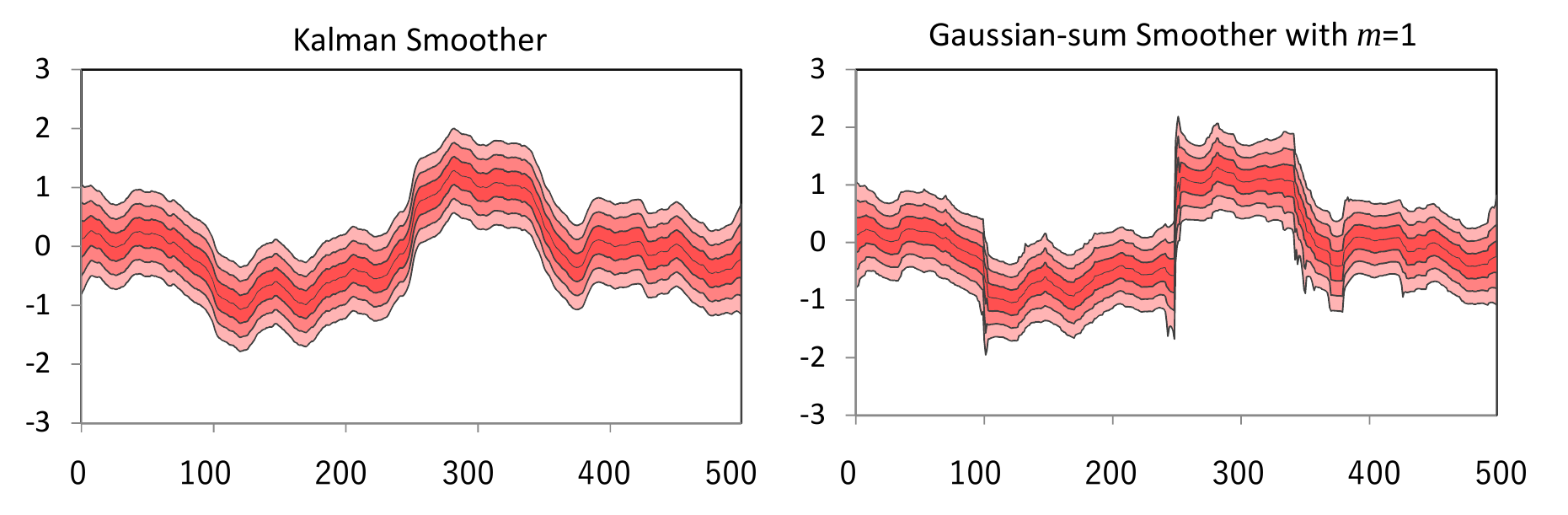}
\end{center}
\caption{Comparison with Kalman smoother and Gaussian-sum smoother with one component ($m$=1). }
\label{Fig_Gaussian-sum-with-m=1}
\vspace{5mm}
\end{figure}

It is interesting to note that as seen in Figure \ref{Fig_Gaussian-sum-with-m=1} 
the Gaussian-sum smoother with $m=1$ 
is different from the Kalman smoother.

\section{Conclusion}

Pearson $\chi^2$-divergence of two Gaussian components with respect to the merged single Gaussian distribution has an explicit analytical form.
According to the empirical studies, sequential reduction method based on the
Pearson $\chi^2$-divergence performed almost similarly as the one
based on the Kullback-Leibler divergence for which computationally costly
numerical integration is necessary.
Application to Gaussian-sum filter and smoother is shown
and it is shown that Gaussian-sum filtering method is very efficient
for linear state-space model with Gaussian mixture noise inputs.

\newpage
\section{Appendix}
\begin{small}
In this appendix, it will be shown that 
\begin{eqnarray}
\int \frac{f_j(x)f_k(x)}{p_{jk}(x)} dx
&=& (2\pi )^{-\frac{k}{2}}\left|\Sigma_j\right|^{-\frac{1}{2}}\left|\Sigma_k\right|^{-\frac{1}{2}}\left| V_{jk}\right|^{\frac{1}{2}} \left| W_{jk}\right|^{-\frac{1}{2}}
\exp\left\{-\frac{1}{2}(\mu_j - \mu_k)^T (\Sigma_j + \Sigma_k)^{-1}(\mu_j - \mu_k)\right\}\nonumber \\
 && \times\exp\left\{ -\frac{1}{2}(\zeta_{jk} -\eta_{jk})^T (V_{jk}-\Sigma_{jk} )^{-1}(\zeta_{jk} -\eta_{jk} )\right\}   \label{eq_32}
\end{eqnarray}
which is used in the derivation of the equation (\ref{eq_integral_of fjfk_over_p}).

\vspace{5mm}
\noindent{\bf Notations}
\begin{eqnarray}
 \Sigma_{jk}^{-1} &=& \Sigma_j^{-1}+\Sigma_k^{-1}, \quad
 \Sigma_{jk} = (\Sigma_{jk}^{-1})^{-1} = (\Sigma_j^{-1}+\Sigma_k^{-1})^{-1}, \\
\xi_{jk} &=& (\alpha_j+\alpha_k)^{-1}(\alpha_j\mu_j+\alpha_k\mu_k) \\
V_{jk} &=& (\alpha_j+\alpha_k)^{-1}
        \left[\alpha_j\left\{\Sigma_j+(\mu_j-\xi_{jk})(\mu_j-\xi_{jk})^T\right\}
            + \alpha_k\left\{\Sigma_k+(\mu_j-\xi_{jk})(\mu_j-\xi_{jk})^T\right\}\right] \\
W_{jk}      &=& \Sigma_j^{-1}+\Sigma_k^{-1}-V_{jk}^{-1} = \Sigma_{jk}^{-1}-V_{jk}^{-1}, \\
\zeta_{jk}  &=& (\Sigma_j^{-1}+\Sigma_k^{-1})^{-1} 
     (\Sigma_j^{-1}\mu_j + \Sigma_k^{-1}\mu_k) \\
\Sigma_{jk}^{-1}\zeta_{jk} &=& \Sigma_j^{-1}\mu_j + \Sigma_k^{-1}\mu_k ,\\
\eta_{jk}   &=& \left(\Sigma_j^{-1}+\Sigma_k^{-1}-V_{jk}^{-1}\right)^{-1}
\left(\Sigma_j^{-1}\mu_j + \Sigma_k^{-1}\mu_k-V_{jk}^{-1}\xi_{jk}\right) , \nonumber \\
    &=& \left(\Sigma_{jk}^{-1} -V_{jk}^{-1}\right)^{-1}
     \left( \Sigma_{jk}^{-1}\zeta_{jk} -V_{jk}^{-1}\xi_{jk} \right),\\
\bar{W}_j &=& 2\Sigma_j^{-1} -V_{jk}^{-1}, \\
\zeta_j &=& (2\Sigma_j^{-1})^{-1}(2\Sigma_j^{-1}\mu_j )= \Sigma_j \Sigma_j^{-1}\mu_j =\mu_j \\ 
\eta_j &=& (2\Sigma_j^{-1}-V_{jk}^{-1})^{-1}(2\Sigma_j^{-1}\mu_j - V_{jk}^{-1}\xi_{jk} ).
\end{eqnarray}
Hereafter in this appendix, for the simplicity of the notation, the suffix ${}_{jk}$ is omitted, 
namely we denote $\xi_{jk}=\xi$, $V_{jk}\equiv V$, $\Sigma_{jk}\equiv \Sigma$, $\zeta_{jk}\equiv \zeta$, $\xi_{jk}\equiv \xi$, $\eta_{jk}=\eta$.

\vspace{5mm}
\noindent{\bf Matrix Lemma}
\begin{eqnarray}
(\Sigma_j^{-1}+\Sigma_k^{-1})^{-1} 
  &=&  \Sigma_j - \Sigma_j (\Sigma_j+\Sigma_k)^{-1}\Sigma_j, \\
(\Sigma^{-1}-V^{-1})^{-1} &=& \Sigma(V-\Sigma )^{-1}V, \\
(\Sigma_j^{-1}+\Sigma_k^{-1}-V^{-1})^{-1}
&=& (\Sigma_j^{-1}+\Sigma_k^{-1})^{-1} (V- (\Sigma_j^{-1}+\Sigma_k^{-1})^{-1})^{-1}V  \nonumber\\
&=& \left\{\Sigma_j - \Sigma_j(\Sigma_j + \Sigma_k)^{-1}\Sigma_j\right\}
    (V- \Sigma )^{-1}V  \\
V^{-1} - V^{-1}\Sigma (V-\Sigma )^{-1} 
  &=& V^{-1}(V-\Sigma )(V-\Sigma )^{-1} - V^{-1}\Sigma (V-\Sigma )^{-1} 
   = (V-\Sigma )^{-1}  \\
\Sigma^{-1} -(V-\Sigma )^{-1} V\Sigma^{-1}
  &=& (V-\Sigma )^{-1}(V-\Sigma )\Sigma^{-1} - (V-\Sigma )^{-1} V\Sigma^{-1}
   = -(V-\Sigma )^{-1} \\
 V^{-1}\Sigma (V-\Sigma )^{-1} V\Sigma^{-1} 
  &=& \{\Sigma V^{-1}(V-\Sigma )\Sigma^{-1}V\}^{-1}  =(V-\Sigma )^{-1}
\end{eqnarray}

\vspace{5mm}
\noindent{\bf Lemma 1}
\begin{eqnarray}
&&\mu_j^T \Sigma_j^{-1} \mu_j + \mu_k^T \Sigma_k^{-1} \mu_k
  -  \left(\Sigma_j^{-1}\mu_j + \Sigma_k^{-1}\mu_k\right)^T
     \left(\Sigma_j^{-1}+\Sigma_k^{-1}\right)^{-1}
     \left(\Sigma_j^{-1}\mu_j + \Sigma_k^{-1}\mu_k\right) \nonumber\\
&& \hspace{20mm}= (\mu_j - \mu_k)^T (\Sigma_j + \Sigma_k)^{-1} (\mu_j -\mu_k)
\end{eqnarray}

\noindent{\bf proof}
\begin{eqnarray}
\lefteqn{ \mu_j^T \Sigma_j^{-1} \mu_j + \mu_k^T \Sigma_k^{-1} \mu_k
  -  \left(\Sigma_j^{-1}\mu_j + \Sigma_k^{-1}\mu_k\right)^T
     \left(\Sigma_j^{-1}+\Sigma_k^{-1}\right)^{-1}
     \left(\Sigma_j^{-1}\mu_j + \Sigma_k^{-1}\mu_k\right)} \nonumber\\
&=& \mu_j^T \Sigma_j^{-1} \mu_j + \mu_k^T \Sigma_k^{-1} \mu_k
  -  \left(\Sigma_j^{-1}\mu_j + \Sigma_k^{-1}\mu_k\right)^T
     \left\{\Sigma_j- \Sigma_j(\Sigma_j+\Sigma_k)^{-1}\Sigma_j\right\}
     \left(\Sigma_j^{-1}\mu_j + \Sigma_k^{-1}\mu_k\right) \nonumber\\
&=& \mu_j^T \Sigma_j^{-1} \mu_j + \mu_k^T \Sigma_k^{-1} \mu_k  \nonumber \\
&&    -\left\{\mu_j^T     -\mu_j^T(\Sigma_j + \Sigma_k)^{-1}\Sigma_j
    + \mu_k^T \Sigma_k^{-T}\Sigma_j 
    -\mu_k^T\Sigma_k^{-1}\Sigma_j(\Sigma_j + \Sigma_k)^{-1}\Sigma_j\right\}
     \left(\Sigma_j^{-1}\mu_j + \Sigma_k^{-1}\mu_k\right) \nonumber\\
&=& \mu_j^T \Sigma_j^{-1} \mu_j + \mu_k^T \Sigma_k^{-1} \mu_k
  - \mu_j^T\Sigma_j^{-1}\mu_j - \mu_j^T\Sigma_k^{-1}\mu_k
  + \mu_j^T(\Sigma_j + \Sigma_k)^{-1}\mu_j
    + \mu_j^T(\Sigma_j + \Sigma_k)^{-1}\Sigma_j\Sigma_k^{-1}\mu_j
\nonumber \\
&&  - \mu_k^T\Sigma_k^{-T}\mu_j - \mu_k^T\Sigma_k\Sigma_j^{-1}\Sigma_k\mu_j 
    + \mu_k^T\Sigma_k^{-1}\Sigma_j(\Sigma_j + \Sigma_k)^{-1}\mu_j
    + \mu_k^T\Sigma_k^{-1}\Sigma_j(\Sigma_j + \Sigma_k)^{-1}\Sigma_j\Sigma_k^{-1}\mu_k \nonumber \\
&=& \mu_k^T \Sigma_k^{-1} \mu_k 
  - \mu_j^T(\Sigma_j + \Sigma_k)^{-1} \mu_k
  - \mu_k^T(\Sigma_j + \Sigma_k)^{-1} \mu_j
  + \mu_j^T(\Sigma_j + \Sigma_k)^{-1} \mu_j
  - \mu_k^T\Sigma_k^{-T}\Sigma_j(\Sigma_j + \Sigma_k)^{-1}\mu_k \nonumber\\
&=& (\mu_j - \mu_k)^T (\Sigma_j + \Sigma_k)^{-1} (\mu_j -\mu_k)
\end{eqnarray}

\vspace{5mm}
\noindent{\bf Lemma 2}
\begin{eqnarray}
\lefteqn{ (x-\mu_j)^T\Sigma_j^{-1} (x-\mu_j) + (x-\mu_k)^T\Sigma_k^{-1} (x-\mu_k)} \nonumber \\
 && \hspace{20mm} = (x - \zeta)^T \Sigma^{-1} (x - \zeta)  
  + (\mu_j - \mu_k)^T (\Sigma_j + \Sigma_k)^{-1} (\mu_j -\mu_k)
\end{eqnarray}

\noindent{\bf Proof}

Using Lemma 1, we have
\begin{eqnarray}
\lefteqn{ (x-\mu_j)^T\Sigma_j^{-1} (x-\mu_j) + (x-\mu_k)^T\Sigma_k^{-1} (x-\mu_k)} \nonumber \\
 &=& x^T \left(\Sigma_j^{-1}+\Sigma_k^{-1}\right)x 
 - x^T \left(\Sigma_j^{-1}\mu_j + \Sigma_k^{-1}\mu_k\right)
 - \left(\Sigma_j^{-1}\mu_j + \Sigma_k^{-1}\mu_k\right)^T x
 + \mu_j^T \Sigma_j^{-1} \mu_j + \mu_k^T \Sigma_k^{-1} \mu_k \nonumber\\
 &=& \big\{x - (\Sigma_j^{-1}+\Sigma_k^{-1})^{-1}
     (\Sigma_j^{-1}\mu_j + \Sigma_k^{-1}\mu_k)\big\}^T
     \/(\Sigma_j^{-1}+\Sigma_k^{-1})
     \big\{x - (\Sigma_j^{-1}+\Sigma_k^{-1})^{-1}
     (\Sigma_j^{-1}\mu_j + \Sigma_k^{-1}\mu_kt)\big\} \nonumber \\
 && + \mu_j^T \Sigma_j^{-1} \mu_j + \mu_k^T \Sigma_k^{-1} \mu_k
  -  (\Sigma_j^{-1}\mu_j + \Sigma_k^{-1}\mu_k)^T
     (\Sigma_j^{-1}+\Sigma_k^{-1})^{-1}
     (\Sigma_j^{-1}\mu_j + \Sigma_k^{-1}\mu_k) \nonumber \\
 &=& (x - \zeta)^T \Sigma^{-1} (x - \zeta)  
  + (\mu_j - \mu_k)^T (\Sigma_j + \Sigma_k)^{-1} (\mu_j -\mu_k)
\end{eqnarray}

\vspace{5mm}
\noindent{\bf Lemma 3}
\begin{eqnarray}
&&\zeta^T \Sigma^{-1} \zeta - \xi^T V^{-1} \xi
  -  \left(\Sigma^{-1}\zeta - V^{-1}\xi\right)^T
     \left(\Sigma^{-1}-V^{-1}\right)^{-1}
     \left(\Sigma^{-1}\zeta - V^{-1}\xi\right) \nonumber\\
&& \hspace{20mm} = -(\zeta - \xi)^T (V-\Sigma )^{-1} (\zeta -\xi)
\end{eqnarray}

\noindent{\bf Proof}
\begin{eqnarray}
\lefteqn{ \zeta^T \Sigma^{-1} \zeta - \xi^T V^{-1} \xi
  -  \left(\Sigma^{-1}\zeta - V^{-1}\xi\right)^T
     \left(\Sigma^{-1}-V^{-1}\right)^{-1}
     \left(\Sigma^{-1}\zeta - V^{-1}\xi\right)} \nonumber\\
&=& \zeta^T \Sigma^{-1} \zeta - \xi^T V^{-1} \xi
  -  \left(\Sigma^{-1}\zeta - V^{-1}\xi\right)^T
     \Sigma (V- \Sigma)^{-1}V
     \left(\Sigma^{-1}\zeta - V^{-1}\xi\right) \nonumber\\
&=& \zeta^T \Sigma^{-1} \zeta - \xi^T V^{-1} \xi
  - (\zeta^T -\xi^tV^{-1}\Sigma )(V-\Sigma )^{-1} (V\Sigma^{-1}\zeta - \xi ) \nonumber\\
&=& \zeta^T \Sigma^{-1} \zeta + \xi^T V^{-1} \xi
  - \zeta^T(V- \Sigma)^{-1}V\Sigma^{-1}\zeta
  - \zeta^T(V-\Sigma )^{-1}\xi \nonumber \\
&& + \xi^T V^{-1}\Sigma(V-\Sigma )^{-1}V\Sigma^{-1}\zeta
  + \xi^T V^{-1}\Sigma(V-\Sigma )^{-1}\xi
\nonumber \\
&=& -(\zeta - \xi)^T (V-\Sigma )^{-1} (\zeta -\xi)
\end{eqnarray}

\vspace{5mm}
\noindent{\bf Lemma 4}
\begin{eqnarray}
(x-\zeta)^T \Sigma^{-1} (x-\zeta) - (x-\xi)^T V^{-1} (x-\xi) 
  = (x - \eta)^T W (x - \eta) 
     - (\zeta - \xi)^T (V-\Sigma )^{-1} (\zeta - \xi) 
\end{eqnarray}

\noindent{\bf Proof}

Using Lemma 3,
\begin{eqnarray}
\lefteqn{ (x-\zeta)^T \Sigma^{-1} (x-\zeta) - (x-\xi)^T V^{-1} (x-\xi) }\nonumber\\
 &=&x^T (\Sigma^{-1}-V^{-1})x 
 - x^T (\Sigma^{-1}\zeta - V^{-1}\xi )
 - (\Sigma^{-1}\zeta - V^{-1}\xi )^T x
 + \zeta^T \Sigma^{-1} \zeta  - \xi^T V^{-1}\xi  \nonumber\\
 &=& \left\{x - \left(\Sigma^{-1} -V^{-1}\right)^{-1}
     \left( \Sigma^{-1}\zeta -V^{-1}\xi \right)\right\}^T
     \left(\Sigma^{-1} - V^{-1}\right) 
     \left\{x - \left( \Sigma^{-1} - V^{-1}\right)^{-1}
     \left( \Sigma^{-1}\zeta - V^{-1}\xi \right)\right\} \nonumber \\
 && + \zeta^T \Sigma^{-1} \zeta  - \xi^T V^{-1}\xi 
  -  \left( \Sigma^{-1}\zeta -V^{-1}\xi \right)^T
     \left(\Sigma^{-1} - V^{-1}\right)^{-1}
     \left( \Sigma^{-1}\zeta-V^{-1}\xi\right) \nonumber \\
 &=& (x - \eta)^T \left(\Sigma^{-1} - V^{-1}\right) (x - \eta)
      + \zeta^T \Sigma^{-1} \zeta  - \xi^T V^{-1}\xi \nonumber \\
 && -  \left( \Sigma^{-1}\zeta -V^{-1}\xi \right)^T
     \left(\Sigma^{-1}-V^{-1}\right)^{-1}
     \left( \Sigma^{-1}\zeta-V^{-1}\xi\right) \nonumber \\
 &=& (x - \eta)^T W (x - \eta) 
     + \zeta^T \Sigma^{-1} \zeta  - \xi^T V^{-1}\xi 
     - \left( \Sigma^{-1}\zeta -V^{-1}\xi \right)^T  W^{-1}
     \left( \Sigma^{-1}\zeta-V^{-1}\xi\right)  \nonumber \\
 &=& (x - \eta)^T W (x - \eta) 
     - (\zeta - \xi)^T (V-\Sigma )^{-1} (\zeta - \xi) 
\end{eqnarray}

\vspace{5mm}
\noindent{\bf Lemma 5}
\begin{eqnarray}
\lefteqn{ \mu_j^T \Sigma_j^{-1} \mu_j + \mu_k^T \Sigma_k^{-1} \mu_k 
  - \xi^T V^{-1} \xi } \nonumber \\
&=& (\mu_j -\mu_k)^T(\Sigma_j + \Sigma_k )^{-1}(\mu_j -\mu_k ) 
   -(\zeta - \xi )^T(V-\Sigma)^{-1}(\zeta - \xi ).
\end{eqnarray}

\noindent{\bf Proof}
\begin{eqnarray}
\lefteqn{ \mu_j^T \Sigma_j^{-1} \mu_j + \mu_k^T \Sigma_k^{-1} \mu_k 
  - \xi^T V^{-1} \xi } \nonumber \\
 && -  \left(\Sigma_j^{-1}\mu_j + \Sigma_k^{-1}\mu_k - V^{-1}\xi\right)^T
     \left(\Sigma_j^{-1}+\Sigma_k^{-1}-V^{-1}\right)^{-1}
     \left(\Sigma_j^{-1}\mu_j+ \Sigma_k^{-1}\mu_k  - V^{-1}\xi\right) 
\nonumber\\
&=& \mu_j^T \Sigma_j^{-1} \mu_j + \mu_k^T \Sigma_k^{-1} \mu_k 
 -  \left(\Sigma_j^{-1}\mu_j + \Sigma_k^{-1}\mu_k\right)^T
     \left(\Sigma_j^{-1}+\Sigma_k^{-1}\right)^{-1}
     \left(\Sigma_j^{-1}\mu_j + \Sigma_k^{-1}\mu_k\right)    \nonumber \\
&&  - \xi^T V^{-1} \xi   +  \left(\Sigma_j^{-1}\mu_j + \Sigma_k^{-1}\mu_k\right)^T
     \left(\Sigma_j^{-1}+\Sigma_k^{-1}\right)^{-1}
     \left(\Sigma_j^{-1}\mu_j + \Sigma_k^{-1}\mu_k\right)\nonumber \\
&&  -  \left(\Sigma_j^{-1}\mu_j + \Sigma_k^{-1}\mu_k - V^{-1}\xi\right)^T
    (\Sigma^{-1} - V^{-1})
     \left(\Sigma_j^{-1}\mu_j + \Sigma_k^{-1}\mu_k - V^{-1}\xi\right) \nonumber\\
&=&  (\mu_j -\mu_k)^T(\Sigma_j + \Sigma_k )^{-1}(\mu_j -\mu_k )   \\
&& - \xi^T V^{-1} \xi +  \left(\Sigma_j^{-1}\mu_j + \Sigma_k^{-1}\mu_k\right)^T
     \left(\Sigma_j^{-1}+\Sigma_k^{-1}\right)^{-1} 
     \left(\Sigma_j^{-1}\mu_j + \Sigma_k^{-1}\mu_k\right)\nonumber \\
&& -  \left(\Sigma_j^{-1}\mu_j + \Sigma_k^{-1}\mu_k - V^{-1}\xi\right)^T
     (\Sigma^{-1} - V^{-1})
     \left(\Sigma_j^{-1}\mu_j + \Sigma_k^{-1}\mu_k - V^{-1}\xi\right) \nonumber
\end{eqnarray}
Here, the terms after the second term of the above equation can be expressed in a signle term as follows:
\begin{eqnarray}
\lefteqn{ - \xi^T V^{-1} \xi 
 +  \left(\Sigma_j^{-1}\mu_j + \Sigma_k^{-1}\mu_k\right)^T
     \left(\Sigma_j^{-1}+\Sigma_k^{-1}\right)^{-1} 
     \left(\Sigma_j^{-1}\mu_j + \Sigma_k^{-1}\mu_k\right) } \nonumber \\
&& -  \left(\Sigma_j^{-1}\mu_j + \Sigma_k^{-1}\mu_k - V^{-1}\xi\right)^T
     (\Sigma^{-1} - V^{-1})
     \left(\Sigma_j^{-1}\mu_j + \Sigma_k^{-1}\mu_k - V^{-1}\xi\right) \nonumber\\
&=& -\xi V^{-1}\xi  + \zeta^T \Sigma^{-1}\zeta
- (\Sigma^{-1}\zeta - V^{-1}\xi )^T (\Sigma^{-1}-V^{-1})(\Sigma^{-1}\zeta - V^{-1}\xi )    \nonumber \\
&=& -\xi V^{-1}\xi    + \zeta^T \Sigma^{-1}\zeta 
- (\Sigma^{-1}\zeta - V^{-1}\xi )^T \Sigma(V-\Sigma)^{-1}V
    (\Sigma^{-1}\zeta - V^{-1}\xi )  \nonumber \\
&=& -\xi V^{-1}\xi   + \zeta^T \Sigma^{-1}\zeta
 - (\zeta^T - \xi^TV^{-1}\Sigma )^T (V-\Sigma)^{-1}
    (V\Sigma^{-1}\zeta - \xi )   \nonumber \\
&=& \zeta^T \Sigma^{-1}\zeta - \xi V^{-1}\xi  - \zeta^T(V-\Sigma)^{-1}V\Sigma^{-1}\zeta
  +  \xi^T V^{-1}\Sigma(V-\Sigma)^{-1}V\Sigma^{-1}\zeta \nonumber \\
   &&  +\zeta^{-1}(V-\Sigma)^{-1}\xi -\xi^T V^{-1}\Sigma(V-\Sigma)^{-1}\xi \nonumber \\
&=& - \zeta^T (V-\Sigma)^{-1}\zeta - \xi (V-\Sigma)^{-1}\xi + \zeta^T(V-\Sigma)^{-1}\xi
  +  \xi^T (V-\Sigma)^{-1}\zeta \nonumber \\
&=& -(\zeta - \xi )^T(V-\Sigma)^{-1}(\zeta - \xi ).
\end{eqnarray}

\vspace{5mm}
\noindent{\bf Proposition}

Assume that $f_j(x)$, $f_j(x)$, $f_k(x)$ and $p_{jk}(x)$
are respectively given by
$f_j(x)\sim N(\mu_j,\Sigma_j)$, $f_k(x)\sim N(\mu_k,\Sigma_k)$
and $p_{jk}(x)\sim N(\zeta,V)$, 
then integral of $f_j(x)f_k(x)/p_{jk}(x)$ over the whole
domain is given by
\begin{eqnarray}
\int \frac{f_j(x)f_k(x)}{p_{jk}(x)} dx
&=& \left|\Sigma_j\right|^{-\frac{1}{2}}\left|\Sigma_k\right|^{-\frac{1}{2}}
  \left| V\right|^{\frac{1}{2}} \left| W\right|^{-\frac{1}{2}} 
\exp\left\{\frac{1}{2}(\zeta - \xi)^T (V-\Sigma )^{-1}(\zeta - \xi)\right\}
\nonumber\\
&& \times \exp\left\{-\frac{1}{2}(\mu_j - \mu_k)^T 
   (\Sigma_j + \Sigma_k)^{-1}(\mu_j - \mu_k)\right\}
\end{eqnarray}

\noindent{\bf Proof}

Since $f_j(x)$ and $f_k(x)$ are defined by
\begin{eqnarray}
f_j(x) &=& (2\pi )^{-\frac{k}{2}}\left|\Sigma_j\right|^{-\frac{1}{2}}
  \exp\left\{ -\frac{1}{2}(x-\mu_j)^T\Sigma_j^{-1} (x-\mu_j)\right\}  \nonumber\\
f_k(x) &=& (2\pi )^{-\frac{k}{2}}\left|\Sigma_k\right|^{-\frac{1}{2}}
   \exp\left\{ -\frac{1}{2}(x-\mu_k)^T\Sigma_k^{-1} (x-\mu_k)\right\} ,
\end{eqnarray}
respectively, $f_j(x)f_k(x)$ is given by%
\begin{eqnarray}
f_j(x)f_k(x) 
= (2\pi )^{-k}\left|\Sigma_j\right|^{-\frac{1}{2}}\left|\Sigma_k\right|^{-\frac{1}{2}}
  \exp\left\{ -\frac{1}{2}(x-\mu_j)^T\Sigma_j^{-1} (x-\mu_j) 
              -\frac{1}{2}(x-\mu_k)^T\Sigma_k^{-1} (x-\mu_k)\right\}. \label{eq_38}
\end{eqnarray}
Then by Lemma 2
\begin{eqnarray}
f_j(x)f_k(x) 
 &=& (2\pi )^{-k}\left|\Sigma_j\right|^{-\frac{1}{2}}\left|\Sigma_k\right|^{-\frac{1}{2}}
  \exp\left\{ -\frac{1}{2}(x-\zeta)^T \Sigma^{-1} (x-\zeta) \right\}
\nonumber \\
 & &\times  \exp\left\{-\frac{1}{2}(\mu_j - \mu_k)^T (\Sigma_j + \Sigma_k)^{-1} (\mu_j -\mu_k)\right\}. \nonumber 
\end{eqnarray}
Since $p_{jk}(x)$ is defined by
\begin{eqnarray}
p_{jk}(x) &=& (2\pi )^{-\frac{k}{2}}\left|V\right|^{-\frac{1}{2}}   
  \exp\left\{ -\frac{1}{2}(x-\xi)^TV^{-1} (x-\xi)\right\} , 
\end{eqnarray}
$f_j(x)f_k(x)/p_{jk}(x)$ is given by
\begin{eqnarray}
\frac{f_j(x)f_k(x)}{p_{jk}(x)}
&=& (2\pi )^{-\frac{k}{2}}\left|\Sigma_j\right|^{-\frac{1}{2}}\left|\Sigma_k\right|^{-\frac{1}{2}}\left| V\right|^{\frac{1}{2}} 
\exp\left\{-\frac{1}{2}(\mu_j - \mu_k)^T (\Sigma_j + \Sigma_k)^{-1}(\mu_j - \mu_k)\right\}\nonumber \\
 && \times\exp\left\{ -\frac{1}{2}(x-\zeta)^T \Sigma^{-1} (x-\zeta) 
              +\frac{1}{2}(x-\xi)^T V^{-1} (x-\xi)   \right\} .
\end{eqnarray}
Then by Lemma 4, it can be expressed as
\begin{eqnarray}
\frac{f_j(x)f_k(x)}{p_{jk}(x)}  
&=& (2\pi )^{-\frac{k}{2}}\left|\Sigma_j\right|^{-\frac{1}{2}}
  \left|\Sigma_k\right|^{-\frac{1}{2}}\left| V\right|^{\frac{1}{2}} 
\exp\left\{-\frac{1}{2}(\mu_j - \mu_k)^T (\Sigma_j + \Sigma_k)^{-1}(\mu_j - \mu_k)\right\}
  \nonumber\\
&& \times
  \exp\left\{\frac{1}{2}(\zeta - \xi)^T (V-\Sigma)^{-1}(\zeta - \xi)\right\}
  \exp\left\{ -\frac{1}{2}(x-\eta)^T W (x-\eta) \right\} . \label{eq-f1f2overp}
\end{eqnarray}
By integrating whole domain of $x$, we obtain
\begin{eqnarray}
\int \frac{f_j(x)f_k(x)}{p_{jk}(x)} dx
&=& \left|\Sigma_j\right|^{-\frac{1}{2}}\left|\Sigma_k\right|^{-\frac{1}{2}}
  \left| V\right|^{\frac{1}{2}} \left| W\right|^{-\frac{1}{2}} 
\exp\left\{\frac{1}{2}(\zeta - \xi)^T (V-\Sigma )^{-1}(\zeta - \xi)\right\}
\nonumber\\
&& \times \exp\left\{-\frac{1}{2}(\mu_j - \mu_k)^T 
   (\Sigma_j + \Sigma_k)^{-1}(\mu_j - \mu_k)\right\},
\end{eqnarray}
which complete the proof of the proposition.\\

By putting $\mu_k = \mu_j$ ,$\Sigma_k = \Sigma_j$ in the Proposition
we obtain the following 

\noindent{\bf Corollary}
\begin{eqnarray}
\int \frac{f(x)_j^2}{p_{jk}(x)} dx
&=& \left|\Sigma_j\right|^{-1}
  \left| V\right|^{\frac{1}{2}} \left| {W}_j \right|^{-\frac{1}{2}} 
\exp\left\{\frac{1}{2}(\mu_j - \xi )^T (V -\frac{1}{2}\Sigma_j)^{-1}(\mu_j - \xi )\right\}.
\end{eqnarray}

\vspace{5mm}
\noindent{\bf Note}

The equation (\ref{eq-f1f2overp}) can be directly obtained by 
considering the expression of the $f_j(x)f_k(x)/p_{jk}(x)$ as follows:
\begin{eqnarray}
\frac{f_j(x)f_k(x)}{p_{jk}(x)}
 &=& (2\pi )^{-\frac{k}{2}}\left|\Sigma_j\right|^{-\frac{1}{2}}\left|\Sigma_k\right|^{-\frac{1}{2}}\left| V\right|^{\frac{1}{2}} \\
 &\times&  \exp\left\{ -\frac{1}{2}(x-\mu_j)^T\Sigma_j^{-1} (x-\mu_j) 
              -\frac{1}{2}(x-\mu_k)^T\Sigma_k^{-1} (x-\mu_k)
              +\frac{1}{2}(x-\xi)^T V^{-1} (x-\xi)
\right\}. \nonumber 
\end{eqnarray}
Here the terms in the brace of the right hand side of the above equation
is given by
\begin{eqnarray}
\lefteqn{ (x-\mu_j)^T\Sigma_j^{-1} (x-\mu_j) + (x-\mu_k)^T\Sigma_k^{-1} (x-\mu_k) - (x-\xi)^T V^{-1} (x-\xi)}
\nonumber \\
 &=& x^T \left(\Sigma_j^{-1}+\Sigma_k^{-1}-V^{-1}\right)x 
 - x^T \left(\Sigma_j^{-1}\mu_j + \Sigma_k^{-1}\mu_k - V^{-1}\xi \right) \nonumber \\
 &&- \left(\Sigma_j^{-1}\mu_j + \Sigma_k^{-1}\mu_k - V^{-1}\xi \right)^T x
 + \mu_j^T \Sigma_j^{-1} \mu_j + \mu_k^T \Sigma_k^{-1} \mu_k  - \xi^T V^{-1}\xi \nonumber\\
 &=& \left(x - \zeta\right)^{-1} W \left(x - \zeta\right)^{-1}
   + \mu_j^T \Sigma_j^{-1} \mu_j + \mu_k^T \Sigma_k^{-1} \mu_k
     - \xi^T V^{-1}\xi \nonumber \\
 && -  \left(\Sigma_j^{-1}\mu_j + \Sigma_k^{-1}\mu_k -V^{-1}\xi\right)^T
     \left(\Sigma_j^{-1}+\Sigma_k^{-1}-V^{-1}\right)^{-1}
     \left(\Sigma_j^{-1}\mu_j + \Sigma_k^{-1}\mu_k-V^{-1}\xi\right).
\end{eqnarray}
Then by Lemma 5, it can be expressed as 
\begin{eqnarray}
\left(x - \zeta\right)^{-1} W \left(x - \zeta\right)^{-1}
   +  (\mu_j - \mu_k)^T (\Sigma_j + \Sigma_k)^{-1} (\mu_j -\mu_k)
- (\zeta -\xi )^T(V-\Sigma )^{-1}(\zeta -\xi).
\end{eqnarray}
Therefore we obtain the equation (\ref{eq-f1f2overp}).

\end{small}

\vspace{15mm}
\noindent{\Large\bf Aknowledgements}

This work was supported in part by JSPS KAKENHI Grant Number 18H03210.


\begin{thebibliography}{2}

\bibitem{AS 1972}
Alspach, D. and Sorenson, H. (1972). Nonlinear Bayesian estimation using Gaussian sum approximations. {\it IEEE transactions on automatic control}, Vol. 17, No.4, 439--448.

\bibitem{CWPS 2011}
Crouse, D. F., Willett, P., Pattipati, K. and Svensson, L. (2011). 
A look at Gaussian mixture reduction algorithms. 
In {\it 14th International Conference on Information Fusion, IEEE}, 1--8.

\bibitem{Kitagawa 1987}
Kitagawa, G. (1987). Non-Gaussian state-space modeling of nonstationary time series. 
{\it Journal of the American Statistical Association}, Vol. 82, No.400, 1032--1041.

\bibitem{Kitagawa 1989}
Kitagawa, G. (1989). Non-Gaussian seasonal adjustment, {\it Computers \& Mathematics with Applications}, 
Vol.18, No.6/7, pp. 503--514.


\bibitem{Kitagawa 1994}
Kitagawa, G. (1994).  The two-filter formula for smoothing and an implementation of the Gaussian-sum smoother,
{\it Annals of the Institute of Statistical Mathematics}, Vol. 46, No.4, pp. 605--623.


\bibitem{Kitagawa 1996} 
Kitagawa, G. (1996). Monte Carlo filter and smoother for non-Gaussian nonlinear state space models, 
{\it Journal of Computational and Graphical Statistics}, Vol.5, no.1, pp. 1--25.


\bibitem{Runnalls 2007}
Runnalls, A.R. (2007). A Kullback-Leibler approach to Gaussian mixture reduction, 
{\it IEEE Trans. Aerospace and Electronics Systems}, Vol. 43, No. 3, pp. 989--999.

\bibitem{Salmond 1990}
Salmond, D.L. (1990). Mixture reduction algorithms for target tracking in clutter, 
in {\it Signal and Data Processing of Small Targets 1990, Proc. of SPIE}, 1305, 434--445.

\bibitem{SA 1971}
Sorenson, H. W. and Alspach, D. L. (1971). Recursive Bayesian estimation using Gaussian sums. {\it Automatica}, Vol. 7, No.4, 465--479.

\bibitem{West 1993}
West, M. (1993). Approximate posterior distributions by mixture,
{\it Journal of the Royal Statistical Society, Series B (Methodological)}, Vol. 55, No. 2,
pp. 409--422.

\bibitem{WM 2003}
Williams, J.L. and Maybeck, P.S. (2003). Cost-function-based Gaussian mixture reduction, 
in {\it Sixth Int. Conf. on Information Fusion}, Vol. 2, pp. 1047--1054, Piscataway, NJ: IEEE Publ.



\end{thebibliography}
\end{document}